\documentclass[aps,prl,twocolumn,showpacs]{revtex4}
\usepackage{graphicx}

\newcommand{\be}{\begin{equation}}
\newcommand{\ee}{\end{equation}}

\begin{document}

\title{Barkhausen Noise and Critical Scaling in the Demagnetization Curve}

\author{John H.\ Carpenter}
%\email{jhcarpen@uiuc.edu}
\author{Karin A.\ Dahmen}
%\email{dahmen@uiuc.edu}
\affiliation{
University of Illinois at Urbana-Champaign,
Department of Physics,
1110 West Green Street, 
Urbana, IL 61801.
}

\date{\today}

\begin{abstract}

The demagnetization curve, or initial magnetization curve, is studied
by examining the embedded Barkhausen noise using the non-equilibrium,
zero temperature random-field Ising model. The demagnetization curve
is found to reflect the critical point seen as the system's disorder
is changed. Critical scaling is found for avalanche sizes and the size
and number of spanning avalanches. The critical exponents are derived
from those related to the saturation loop and subloops. Finally, the
behavior in the presence of long range demagnetizing fields is
discussed. Results are presented for simulations of up to one million
spins.

\end{abstract}

\pacs{75.60.Ej, 64.60.Ht, 75.60.Ch}

\maketitle

%%%%%%%%%%%%%%%%%%%%%%%%%%%%%%%%%%%%%%%%%%%%%%%%%%%%%%%%%%%%%%%%%%%%%%%%%%%%%%%

%\section{Introduction}

During the last decade, much progress has been made in the study of
universality in hysteresis and avalanche type noise in both experiments
and theory. Hysteresis in ferromagnets and associated Barkhausen noise
have been studied in great detail as a conveniently accessible example
system.
The zero temperature non-equilibrium random field Ising model (RFIM)
and recent variants have proven very successful in predicting 
universal scaling behavior in such systems
\cite{Dahmen-Sethna-PRB96,Kuntz-Sethna,nature}. In particular,
an underlying disorder induced non-equilibrium phase transition has
been found, which may be the cause of the broad range of power law
scaling observed in many systems with avalanche-type noise. It is expected
to be relevant to Barkhausen noise in disordered (hard) magnets 
\cite{hysterI}. For soft magnets, where
the long range demagnetizing fields are important, the system
appears to naturally operate at the single domain wall 
depinning point \cite{nature,Zap98,Zap98b}, thus being self organized critical
\cite{Bak}.

\begin{figure}
\includegraphics[width=8.6cm]{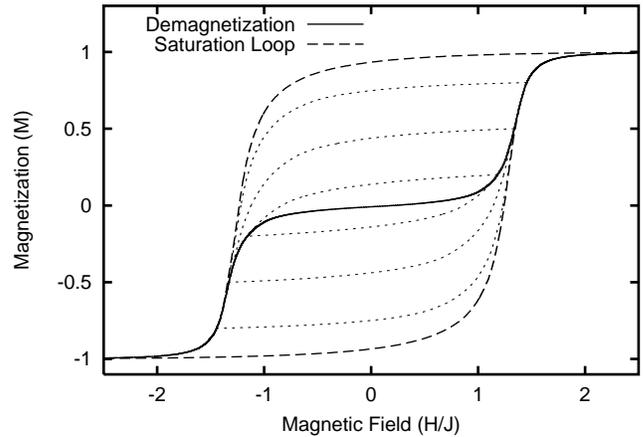}
\caption[Saturation Loop and Demagnetization Curve] { 
\label{rmh-fig}
The saturation hysteresis loop (dashed line) with its corresponding
demagnetization curve (solid line) at a disorder of $R = 2.9J$ in a
$100^3$ system. A few of the many subloops performed are shown as
dotted lines.
}
\end{figure}

So far, most of the theoretical studies of this scaling behavior 
have focused on the properties of the saturation hysteresis loop
(except for some recent 
studies \cite{Car01,Carnp,Zapperi1,Zapperi2,Zapperi3}).
In this letter we report history induced scaling behavior
for subloops and in particular the demagnetization curve.
Ferromagnetic materials with a remnant magnetization
can be demagnetized by applying an oscillating magnetic field 
with amplitude slowly decreasing from a large initial value
to zero. Sometimes the final state is also termed AC-demagnetized
state. The oscillating external field with decreasing amplitude
takes the system through concentric subloops (Fig. \ref{rmh-fig}).
The line connecting the tips of the subloops is known as the 
{\it normal or initial magnetization curve}, or {\it demagnetization
curve}. It is to be distinguished from the {\it virgin curve} which
is obtained by thermal demagnetization \cite{Bertotti-book}.
We used the RFIM to simulate the demagnetization curve for 
different amounts of disorder, especially near the critical disorder 
mentioned above. It turns out that this curve
reflects much of the scaling behavior found for the saturation
hysteresis loop. This may be surprising, since the meta-stable states
encountered in the demagnetization curve are completely different from
those of the saturation loop. In the following we briefly describe
the model, then derive the necessary exponent relations, discuss the 
numerical results, and their relevance to experiments.

%\section{Model}
%\label{sec:model}

In the RFIM the magnetic domains of a real magnet are modeled
by spins $s_i=\pm 1$ arranged on a hypercubic lattice. Nearest neighbors
interact ferromagnetically through a positive exchange interaction $J$.
Variants of the model for soft magnets include  
additional long-range interactions due to the demagnetizing field and 
the dipole-dipole interactions. All spins are coupled to a homogeneous
external magnetic field $H(t)$ that is varied adiabatically slowly.
To model disorder in the system, a quenched random magnetic field $h_i$
is added at each site, that is taken from a Gaussian probability 
distribution  $P(h_i) = exp(-h^2_i/2R^2)/\sqrt{2\pi}R$, whose width $R$
is a measure of the amount of disorder in the system.
We write the effective Hamiltonian for a system with $N$ 
spins \cite{Kuntz-Sethna}:
\be
{\mathcal{H}} = -J\sum_{\langle ij\rangle} s_i s_j - \sum_{i}(H(t) + h_i - 
J_{inf} M )s_i,
\label{hamiltonian}
\ee
where $J_{inf}$ is an infinite range demagnetizing field 
, $M= \sum_{i} s_i/N$
is the magnetization of the system,
and $\langle ij\rangle$ stands for nearest neighbor
pairs of spins. Initially (at time $t=-\infty$), $H(-\infty) = -\infty$ and
all the spins are pointing down. Each spin is always aligned with its local 
effective field 
$h_i^{eff} = J\sum_{\langle ij\rangle}s_j + H(t) + h_{i} - J_{inf}M$.  
The external field
$H(t)$ is slowly increased from $-\infty$ until the local field, $h_i^{eff}$, 
of any spin $s_i$ 
changes sign, causing the spin to flip \cite{hysterI}. Each spin flip 
can trigger coupled spins to flip as well, thus leading to an avalanche of
spin flips. During such an avalanche the external magnetic field is kept 
fixed (adiabatic case). It is only changed in between successive avalanches,
to trigger the next spin flip that seeds a new avalanche. The
simulations are based on code both available on the web \cite{webcode} and that
obtained via correspondence \cite{Sethnacode} which has been modified to allow for
subloops in the history. In the following we set $J_{inf} = 0$.

%\section{Demagnetization}
%\label{sec:demag}

A similar process to the AC demagnetization of the laboratory is used in simulating the
demagnetization curve. For a particular realization of random fields,
the system is started at saturation (all spins aligned) and the field
is then oscillated back and forth while being slowly reduced in
magnitude. For an exact demagnetization, the field reversal points are
sufficiently reduced so that the last avalanche of spins does not align
with the field. One may also perform a non-exact demagnetization by
reducing the field by a fixed amount $\Delta H$ each oscillation.

%\section{Exponent Relations and Scaling}
%\label{sec:exprel}

Such a field history produces an infinite set of
nested subloops. The demagnetization curve itself is then composed of the last
avalanche of each subloop, which brings the field back to
its return point. By examining the scaling behavior of these subloops
one should then be able to determine the corresponding behavior for
the demagnetization curve. 

The scaling behavior of subloops has been shown to be dependent on that
of the saturation loop plus one new exponent \cite{Car01,Carnp}. The scaling
forms for the saturation loop and subloops may be combined to give a
general form dependent on the field $H$, disorder $R$, subloop return
point magnetization $M_{max}$, and the parameter of interest. In particular, near
the critical point the
avalanche size distribution scales as
\be 
D(S,H,R,M_{max}) \sim S^{-\tau}{\mathcal
  D}(S^{\sigma\beta\delta}h',S^\sigma r,S^{\sigma'}\epsilon), 
\ee 
the correlation function as
\be 
G(x,H,R,M_{max}) \sim x^{-1/(d-2+\eta)} {\mathcal
  G}(x/\xi(r,h',\epsilon)) 
\ee 
where $\xi(r,h',\epsilon) \sim r^{-\nu} {\mathcal
  Y}(h'/r^{\beta\delta}, r^{-\nu/\nu'}\epsilon)$, and the number of
spanning avalanches scales as
\be 
N(L,H,R,M_{max}) \sim L^{\theta+\beta\delta/\nu}{\mathcal
  N}(L^{1/\nu}r,L^{1/\nu'}\epsilon,h'/r^{\beta\delta}).  
\ee 
Here $h'\equiv h +Br+B_h\epsilon$ is the tilted scaling axis, where $B$
and $B_h$ are non-universal constants, with $h=H-H_c^d$, $r = (R-R_c)/R$, and
$\epsilon = (M_{max_c}-M_{max})/M_{max}$ \cite{Car01,Carnp,Per99,Per95}. Note that
as the last avalanche in a subloop occurs at $M_{max}$, $M_{max}$ also lies on
the demagnetization curve so that $M_{max_c}=M_c^d$. The critical
point of the demagnetization curve is at $(H_c^d,M_c^d)$. Corrections to $r$ and $\epsilon$ are not
significant for the following calculations.

For $\epsilon > 0$ subloops are steepest at their
endpoints, where $h'=0$. From this one may construct the scaling forms
for the demagnetization curve by realizing that at $h'=0$ the
forms for subloops and the demagnetization curve must be identical. As an example, the avalanche size distribution
would give $D_d(S,H,R)=D(S,H,R,M_{max})$ at $h'=0$ or equivalently
$S^{-\tau_d}{\mathcal D}_d(S^{\sigma_d\beta_d\delta_d}h'_d,S^{\sigma_d}r) =
S^{-\tau}{\mathcal D}(0,S^{\sigma}r,S^{\sigma'}\epsilon)$ where
subscript $d$ denotes quantities associated with the demagnetization
curve and $h'_d=h+B_dr$.  Notice that
$h'=0$ implies $\epsilon = -(h+Br)/B_h$ so that both sides have the same
functional dependence. From this we may read off that one expects
$\tau_d = \tau$, $B_d = B$, $\sigma_d\beta_d\delta_d=\sigma'$, and
$\sigma_d = \sigma$.

The integrated avalanche size distribution for the demagnetization
curve is obtained by integrating
over the field $h'_d$ which is equivalent to integrating over
$\epsilon$ at $h'=0$. Performing this integration yields 
\be
D^{int}_d(S,R) \sim
S^{-(\tau_d+\sigma_d\beta_d\delta_d)}{\mathcal D}_d^{int}(S^{\sigma_d}r)
\sim S^{-(\tau+\sigma')}{\mathcal D}^{int}(S^{\sigma}r).
\label{demagaheq}
\ee
From this one
obtains another scaling relation,
$\tau_d+\sigma_d\beta_d\delta_d = \tau + \sigma'$.  

A similar analysis
of the correlation function yields the relations $\nu_d = \nu$,
$\beta_d/\nu_d = (\beta-\beta\delta+\sigma'/\sigma)/\nu$, and $\eta_d
= \eta$.  Also, integrating the number of spanning avalanches over
$\epsilon$ yields the additional relation $\theta_d = \theta
+(\beta\delta-\sigma'/\sigma)/\nu$. The associated scaling forms are
\be
G_d^{int}(x,R) \sim x^{-(d+\beta_d/\nu_d)} {\mathcal G}_d^{int}(x r^{\nu_d}) 
\label{demagcfeq}
\ee
and
\be
N_d^{int}(L,R) \sim L^{\theta_d}{\mathcal N}_d^{int}(L^{1/\nu_d}r)
\label{demagsaeq}
\ee
respectively.
It is possible to determine other relations, but from these we have
already determined all the demagnetization exponents from those
associated with the saturation loop and subloops. A summary of these
relations is given in Table \ref{reltab}.
\begin{table}
\caption[Exponent Relations] {
\label{reltab}
Relations between exponents describing the saturation loop (unprimed),
subloops (primed), and demagnetization curve (subscript $d$) scaling behaviors.
}
\begin{ruledtabular}
\begin{tabular}{c c}

Relation & Source \\
\hline
$\tau_d+\sigma_d\beta_d\delta_d = \tau+\sigma\beta\delta - \sigma\nu X'$ & $D_{int}(S,R)$ \\
$\beta_d/\nu_d = \beta/\nu-X'$ & $G_{int}(x,R)$ \\
$\sigma_d = \sigma$ & $D_{int}(S,R)$ \\
$\theta_d = \theta + X'$ & $N(L,R)$ \\
$\tau_d = \tau$ & $D(S,R,H)$ \\
$\nu_d = \nu$ & $G_{int}(x,R)$ \\
$X' = (\beta\delta-\sigma'/\sigma)/\nu =\footnote{From relations
  1, 3, and 5.}
(\beta\delta-\beta_d\delta_d)/\nu$ &

\end{tabular}
\end{ruledtabular}
\end{table}
Note that all
demagnetization quantities may be expressed as the corresponding
saturation loop value plus a function of the saturation loop
exponents times a correction factor $X' \equiv (\beta\delta -
\sigma'/\sigma)/\nu = \beta\delta/\nu - \beta_d\delta_d/\nu_d$ (see Table \ref{reltab}).

%\section{Simulation Results}
%\label{sec:results}

\begin{figure}
\includegraphics[width=8.6cm]{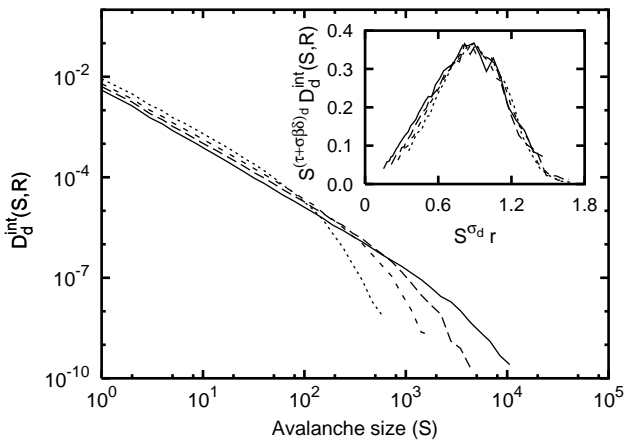}
\caption[Avalanche Distribution] { 
\label{aah-fig}
Integrated avalanche size distribution curves for $100^3$ systems at
disorders $R = 2.5J$, $2.7J$, $2.9J$, and $3.3J$ and averaged over $10$,
$5$, $4$, and $1$ random seeds respectively. The scaling collapse is
shown in the inset where $r = (R-R_c)/R$ with $R_c = 2.12J$. The
critical exponents are $(\tau+\sigma\beta\delta)_d = 2.1$ and
$1/\sigma_d = 4.1$.
}
\end{figure}

The integrated avalanche size distribution was measured on the
demagnetization curve in systems with $100^3$ spins for several
disorders above the critical disorder. The resulting distributions are
shown in Figure \ref{aah-fig}.
As with the saturation loop, the cutoff
size is found to increase as the disorder approaches the critical
disorder. The avalanche distributions were collapsed using the scaling
ansatz given by Eq.~\ref{demagaheq} and are shown in the inset of Fig.~\ref{aah-fig}.  
Within errorbars, the
resulting scaling function is identical in form to that found for the
saturation loop multiplied by a constant.  The resulting exponents are listed in Table
\ref{exptab} and are consistent with the
relations (Table \ref{reltab}) predicted from the previous analysis.
\begin{table}
\caption[Critical Exponents] {
\label{exptab}
Universal critical exponents found from scaling collapses for both the
demagnetization curve and saturation hysteresis loop.
}
\begin{ruledtabular}
\begin{tabular}{c c c}

Exponent & Demagnetization & Saturation\footnote{Reference \cite{Per99,Per95}} \\
\hline
$\tau+\sigma\beta\delta$ & $ 2.10 \pm 0.05 $ & $2.03 \pm 0.03$   \\
$1/\sigma$               & $ 3.9 \pm 0.4 $   & $4.2 \pm 0.3$     \\
$d + \beta/\nu$          & $ 3.1 \pm 0.2 $   & $3.07 \pm 0.30$   \\
$1/\nu$                  & $ 0.71 \pm 0.10 $ & $0.71 \pm 0.09$   \\
$\theta$                 & $ 0.01 \pm 0.01 $ & $0.015 \pm 0.015$ \\
$\beta/\nu$              & $ 0.03 \pm 0.02 $ & $0.025 \pm 0.020$

\end{tabular}
\end{ruledtabular}
\end{table}
Spin-flip correlation functions also collapse according to
Eq.~\ref{demagcfeq} \cite{Carnp}. The measured exponents are listed in Table
\ref{exptab}.

\begin{figure}
\includegraphics[width=8.6cm]{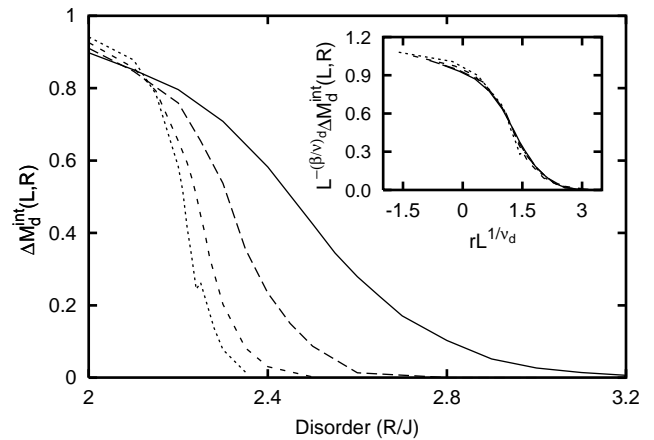}
\caption[Magnetic Discontinuity] { 
\label{fss-fig}
Size of the magnetic discontinuity $\Delta M_d^{int}$ due to system
spanning avalanches for system sizes of $20^3$, $40^3$, $70^3$, and
$100^3$ averaged over $400$, $60$, $80$, and $10$ random
configurations respectively. Curves are collapsed using $r =
(R-R_c)/R$ with $R_c = 2.12J$.  
The inset shows the collapse using the critical 
exponents $\beta/\nu = 0.03$ and $1/\nu = 0.71$.
}
\end{figure}

Although $\beta_d/\nu_d$ was measured in the correlation
function collapse, due to its small magnitude the value cannot be
reliably extracted from the dimension of the system in the equation
$d+\beta_d/\nu_d$. However, one may directly measure $\beta_d/\nu_d$
from $\Delta M^{int}_d$, the change in magnetization due to
all system spanning avalanches \cite{Per99}. For the demagnetization curve, the
scaling form is 
\be
\Delta M_d^{int} \sim L^{-\beta_d/\nu_d}\Delta{\mathcal
  M}_d^{int}(L^{1/\nu_d}r).
\label{demagdmeq}
\ee
The jump in magnetization was measured 
for system sizes of $20^3$, $40^3$, $70^3$, and $100^3$ for
a range of disorders. The resulting curves are shown in
Fig.~\ref{fss-fig} with a scaling collapse using Eq.~\ref{demagdmeq} in
the corresponding inset.
The exponents are listed in Table \ref{exptab}. 

The number of system spanning avalanches in the demagnetization curve
was also measured for the same systems as used in the size of spanning
avalanches collapse. The resulting exponents are listed in Table
\ref{exptab}.  Due to fluctuations the curves do not collapse well
near their peak values. The exponent $\theta_d$, which scales the peak
value, has thus been left with a large uncertainty and could possibly
vanish \cite{Carnp}.

Comparing the exponents from the two finite size scaling collapses
shows that the values of $1/\nu$ are in excellent agreement with the
prediction that they should be equal in the saturation loop and
demagnetization curve. Similarly, the two exponents governing the
power-law behavior at the critical point are found to differ from the
saturation loop values, as predicted. However the error bars have
significant overlap. A simpler relation may be obtained by adding the
equations for $\theta_d$ and $\beta_d/\nu_d$ giving $\theta_d +
\beta_d/\nu_d = \theta + \beta/\nu$ which is satisfied by the measured
exponents. The value of the correction factor may be estimated from
the finite size scaling exponents giving $X' = -0.005 \pm 0.005$. The
power law exponent from the avalanche size distribution also contains
the correction factor, but due to relatively small system sizes,
finite size effects do not allow one to reliably extract $X'$. While
$X'$ was found to be slightly negative, because of the large
uncertainty in the critical exponents for both the saturation loop and
demagnetization curve, one cannot rule out the possibility that $X'$
vanishes (as occurs on the Bethe lattice \footnote{Reference \cite{Zapperi3} finds $\beta_d=\beta$ for the RFIM on the Bethe lattice. The exponent relations of Table \ref{reltab} give $X'=0$.}). If such
were the case then all the critical exponents would be identical for
both the saturation loop and demagnetization curve.

%\section{Long Range}
%\label{sec:longrange}

\begin{figure}
\includegraphics[width=8.6cm]{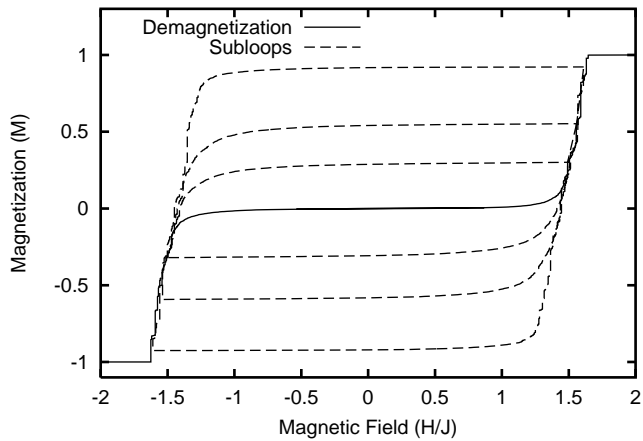}
\caption[Long Range Field Demagnetization Curve] {
\label{lrf-fig}  
The demagnetization curve (solid line) and several subloops (dashed
lines) at a disorder of $R = 1.8J$ and $J_{inf}=0.25J$ in a $100^3$ system.
}
\end{figure}

In the above analysis we have ignored long-range demagnetizing
fields. To examine the effect of such fields one now sets $J_{inf}>0$
in the Hamiltonian, Eq.~\ref{hamiltonian}. A demagnetization process
may again be applied to the system by applying an oscillating field
and decreasing it by a given $\Delta H$ each oscillation. The
resulting curves for a $100^3$ system at a disorder of $R=1.8J$ with
$J_{inf} = 0.25J$ are shown in Fig.~\ref{lrf-fig}. 
For clarity only a
few subloops are shown. All the curves, including the demagnetization
curve, show a linear $M(H)$ behavior as the field is increased, as one
would expect for a system undergoing domain wall propagation.
Additionally, for the same curves one finds a power law distribution
of avalanche sizes, as expected for a self organized critical system
\cite{Zap98,Zap98b}. Preliminary experimental results by A.\ Mills and M.\
B.\ Weissman using a soft magnetic material suggest the results of
Fig.~\ref{lrf-fig} are correct \cite{Mills}.

%\section{Conclusions}
%\label{sec:conclusions}

In the absence of long range demagnetizing fields, the demagnetization
curve of the RFIM exhibits disorder induced critical scaling in quantities associated
with its Barkhausen noise.  Additionally there are no new independent
critical exponents, rather all may be derived from those of the
saturation loop and subloops. Experiments to confirm such behavior
have not yet been performed, but hard magnets or systems where
disorder dominates are expected to exhibit such behavior \cite{Ber00}. Soft magnets
are not expected to exhibit such behavior as realized when long range
demagnetizing fields are included in the RFIM. The system is then
self organized critical, with power law distributions of
avalanches for both the saturation loop and subloops.

\begin{acknowledgments}
We thank Jim Sethna, Andy Mills, and Mike Weissman, for very useful discussions. This work was supported by
NSF grants DMR 99-76550 and DMR 00-72783. K.\ D.\ also gratefully
acknowledges support from an A.\ P.\ Sloan fellowship. We also thank IBM for a
generous equipment award and the Materials Computation Center for the
use of their IBM Workstation Cluster.
\end{acknowledgments}

\end{document}